\documentclass[12pt]{revtex4}
\usepackage{emlines2}
\usepackage[active]{srcltx}
\usepackage{epsfig}
\usepackage{graphicx}
\newcommand{\OO}{\mbox{\boldmath $\Omega$}}

\newcommand{\m}{\mbox{\boldmath $m$}}

\newcommand{\n}{\mbox{\boldmath $n$}}

\newcommand{\ta}{\mbox{\boldmath $\tau$}}
\newcommand{\ro}{\mbox{\boldmath $\rho$}}
\newcommand{\rr}{\mbox{\boldmath $r$}}

\newcommand{\q}{\mbox{\boldmath $q$}}

\newcommand{\p}{\mbox{\boldmath $p$}}

\newcommand{\kk}{\mbox{\boldmath $k$}}
\newcommand{\sn}{\mbox{\scriptsize\boldmath$n$}}

\newcommand{\sm}{\mbox{\scriptsize\boldmath$m$}}
\newcommand{\sta}{\mbox{\scriptsize\boldmath$\tau$}}

\newcommand\lt{\left}
\newcommand\rt{\right}

\newcommand\ola{\overleftarrow}
\newcommand\ora{\overrightarrow}

\newcommand\fr{\displaystyle\frac}
\newcommand{\htts}{\mbox{\boldmath$\hat{t}\kern1pt$}}

\newcommand{\eref}[1]{Eq. (\ref{#1})}
\newcommand\Sh{Schr\"odinger equation}
\begin{document}
\draft
\title{\bf {Scattering of scalar waves on a single crystalline plane.}}
\author{V.K. Ignatovich}
\affiliation{Joint Institute for Nuclear Research}
\address{FLNP JINR 6 Joliot-Curie, Dubna, Moscow region, 141980 Russia, e-mail:v.ignatovi@gmail.com}
\begin{abstract}
Scattering of a scalar particle on a crystalline plane with
quadratic cell and identical fixed scatterers is solved precisely.
Contradiction of the standard scattering theory is pointed out.
\end{abstract}

\maketitle
PACS: 03.65.-w; 25.40.Fq; 61.05.fm;

\section{Introduction}

The standard dynamical diffraction theory was formulated many
years ago (see, for instance~\cite{ew,bac}) and since then enter
all the textbooks almost without a change as is reflected, for
instance in~\cite{bush,hal}. I also did not like this theory as
the Anonymous author of the epigraph and decided to reformulate it
to see clearly all the physical processes that take place in the
diffraction. After many years of lecturing on ``Neutron optics''
and many publications on this topic (see for instance \cite{uig})
I have understood that I reached a perfect level of understanding
to expose some material for teachers of high school in clear and
pedagogical manner. So I decided to submit a single part of this
topic to Am.J.Phys. I should say that during my study of the
subject I found some contradictions in present day scattering
theory~\cite{conc1}  and decided to disclose them on these pages
in hope to initiate some thinking and discussion. The standard
textbooks do not provide food for thinking. They provide only
acquired knowledge that must be digested. And nobody does care
whether the stuff is really digestible. A student who was feeded
by the standard textbooks can become a scientist only, if he saved
his curiosity, is brave enough to permit himself to doubt the
acquired knowledge, and at the same time has a solid mathematical
education. I hope the readers will agree with me.

Here I present my understanding of diffraction of an incident
plane wave $\exp(i\kk_0\rr)$ of a scalar particle on a single
infinite crystalline plane perpendicular to $z$-axis, which
crosses it at the point $z=0$. The plane is inhabited by identical
atoms firmly fixed at their positions and arranged in square
lattice with an elementary cell having sides equal to $a$ and
directed along $x$ and $y$ axes, so that coordinates of atoms are
at $\rr_n=a\n$, where the vector $\n=(n_x,n_y,0)$ has integer
components $n_x$ and $n_y$. A single scatters scatters as a point,
and its scattering is characterized by the scattering amplitude
$b$. So the problem is: how to find precisely diffraction of the
incident wave on such a crystalline plane. This problem can be
considered as a pure quantum or wave mechanical, however it has
relation to real life because the plane wave can be imagined as a
thermal neutron with wave length of 1 \AA, and crystalline plane
as a plane inside a real crystal with $a\approx1$ \AA, and fixed
atoms as real heavy atoms at zero temperature with scattering
amplitude $b\sim10^{-12}$ cm.

We start our story with description of wave function for
scattering of the plane wave on a single atom taken alone. This
wave function, as declared in all textbooks, is\footnote{It is
common to write sign $+$ before scattered spherical wave. I insist
that it is more appropriate to write there minus, which means
redefinition of the sign of the scattering amplitude $b$. With my
choice positive $b$ corresponds to positive optical potential of a
medium $u=4\pi N_0b$, where $N_0$ is atomic density of the medium.
See~\cite{uig}.}
\begin{equation}\label{n1}
\psi(\rr)=\exp(i\kk_0\rr)-\fr br\exp(ikr),
\end{equation}
where $b$ is the scattering amplitude and $k=|\kk_0|$. To find
diffraction on a whole plane we must take into account multiple
wave scattering~\cite{fold,lax,lax1} between atoms in the plane.

\section{Multiple wave scattering}

In \eref{n1} the scattering atom is fixed at the point $r=0$. If
it is fixed at another point $\rr_1$ the wave function looks
\begin{equation}
\psi(\rr)=\exp(i\kk_0\rr)-\exp(i\kk_0\rr_1)\fr{b}{|\rr-\rr_1|}\exp(ik|\rr-\rr_1|)=
\psi_0(\rr)-\psi_0(\rr_1)\fr{b}{|\rr-\rr_1|}\exp(ik|\rr-\rr_1|),
\label{n2}\end{equation} where $\psi_0(\rr)=\exp(i\kk_0\rr)$, and
the factor $\psi_0(\rr_1)$ accounts for the field illuminating the
scatterer.

If we have two scatterers at the points $\rr_{1,2}$, then the
total wave function becomes
\begin{equation}
\psi(\rr)=\exp(i\kk_0\rr)
-\psi_1\fr{b_1}{|\rr-\rr_1|}\exp(ik|\rr-\rr_1|)
-\psi_2\fr{b_2}{|\rr-\rr_2|}\exp(ik|\rr-\rr_2|),
\label{n3}\end{equation} where factors $\psi_{1,2}$ should take
into account rescattering between centers. This rescattering leads
to equations
\begin{equation}
\psi_1=\exp(i\kk_0\rr_1) -\psi_2b_2\eta,\quad
\psi_2=\exp(i\kk_0\rr_2) -\psi_1b_1\eta, \label{n4}\end{equation}
where $\eta=\exp(ik\rho)/\rho$, and $\rho=|\ro|=|\rr_1-\rr_2|$.
Solution of this system of equations is
\begin{equation}
\psi_{1,2}=\fr{\exp(i\kk_0\rr_{1,2})-b_{2,1}\eta\exp(i\kk_0\rr_{2,1})}
{1-b_1b_2\eta^2}=\exp(i\kk_0\rr_{1,2})\fr{1-b_{2,1}\eta\exp(\mp
i\kk_0\ro)}{1-b_1b_2\eta^2}. \label{n5}\end{equation}

The wave function \eref{n3} at large $r$ can be approximated as
\begin{equation}\label{n6}
\psi(\rr)\approx\exp(i\kk_0\rr)-B\fr{\exp(ikr)}{r}.
\end{equation}
Here we used approximation $|\rr-\rr_i|\approx
r-(\rr\cdot\rr_i)/r$, and introduced the total scattering
amplitude of both scatterers
\begin{equation}\label{n7}
B=\psi_1b_1\exp(-i\kk\rr_1)+\psi_2b_2\exp(-i\kk\rr_2),
\end{equation}
where $\kk=k\rr/r$. Substitution of \eref{n5} into \eref{n7} gives
\begin{equation}\label{n8}
B=\fr{b_1\exp(i\q\rr_1)+b_2\exp(i\q\rr_2)-b_1b_2\eta[\exp(i\q\rr_1-i\kk_0\ro)+\exp(i\q\rr_2+i\kk_0\ro)]}{1-b_1b_2\eta^2},
\end{equation}
where $\q=\kk_0-\kk$ is the momentum transferred. From such a
simple exercise one can see that scattering amplitude depends not
solely on momentum transferred $\q$, but also on the other:
incident, $\kk_0$, or scattered momentum $\kk$. Let's note the
denominator in \eref{n8}. At some $\rho$ and $k$ it can be small,
and $|B|$ can be much larger than $b_{1,2}$. There is a temptation
to explain all the nuclear forces in this way.

The above simple exercise is easy to generalize to many
scatterers. The generalization is called multiple wave scattering
(MWS) theory. If we have $N$ fixed scatterers with different
amplitudes $b_i$, then the total wave function with account of
scattering is
\begin{equation}
\psi(\rr)=\exp(i\kk_0\rr)
-\sum_{n=1}^{N}\psi_n\fr{b_n}{|\rr-\rr_n|}\exp(ik|\rr-\rr_n|).
\label{n7a}\end{equation} The field $\psi_j$ illuminating $j$-th
scatterer is determined from the equation
\begin{equation}
\psi_n=\exp(i\kk_0\rr_n) -\sum_{n'\ne n}\psi_{n'}b_{n'}\eta_{n'n},
\label{n8f}\end{equation} where
$\eta_{n'n}=\exp(ik|\rr_n'-\rr_n|)/|\rr_n'-\rr_n|$. There are some
sets of atoms for which the system \eref{n8f} can be easily
solved. One of them is a crystalline plane shown in fig.\ref{nf1}
with identical $b_n=b$ and $\rr_{n}=a\n$, where vector
$\n=(n_x,n_y)$ has integer components $n_{x,y}$.

From symmetry considerations it follows that
\begin{equation}\label{n8a}
\psi_n=C\exp(i\kk_0\rr_n).
\end{equation}
Substitution of this expression into \eref{n8f} gives $C=1-CbS$,
and $C=1/(1+bS)$, where
\begin{equation}
S=\sum_{\sn\ne0}\exp(i(\kk_0\cdot\n) a)\fr{\exp(ika|\n|)}{a|\n|}.
\label{n9}\end{equation}
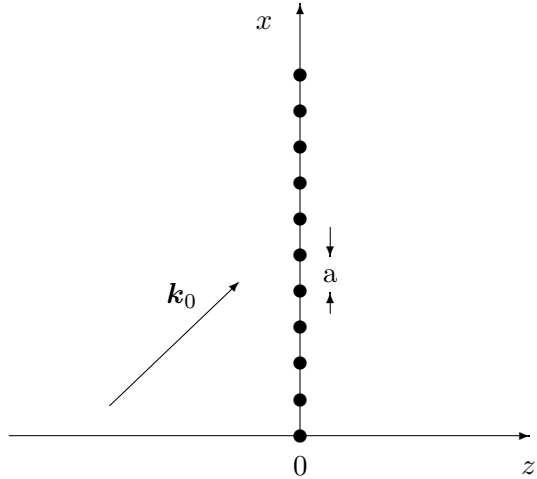
\begin{figure}[thb]
\special{em:linewidth 0.4pt} \unitlength 0.80mm
\linethickness{0.4pt}
\begin{center}
\begin{picture}(106.00,82.00)
\put(106.00,10.33){\vector(1,0){0.2}}
\emline{19.67}{10.33}{1}{106.00}{10.33}{2}
\put(68.00,10.33){\circle*{2.40}}
\put(68.00,16.33){\circle*{2.40}}
\put(68.00,22.33){\circle*{2.40}}
\put(68.00,28.33){\circle*{2.40}}
\put(68.00,34.33){\circle*{2.40}}
\put(68.00,40.33){\circle*{2.40}}
\put(68.00,46.33){\circle*{2.40}}
\put(68.00,52.33){\circle*{2.40}}
\put(68.00,58.33){\circle*{2.40}}
\put(68.00,64.33){\circle*{2.40}}
\put(68.00,70.33){\circle*{2.40}}
\put(57.67,35.67){\vector(1,1){0.2}}
\emline{36.33}{15.33}{3}{57.67}{35.67}{4}
\put(48.33,33.67){\makebox(0,0)[cc]{$\kk_0$}}
\put(68.00,82.00){\vector(0,1){0.2}}
\emline{68.00}{10.33}{5}{68.00}{82.00}{6}
\put(106.00,5.00){\makebox(0,0)[cc]{$z$}}
\put(68.00,5.33){\makebox(0,0)[cc]{0}}
\put(62.00,79.00){\makebox(0,0)[cc]{$x$}}
\put(73.00,37.00){\makebox(0,0)[cc]{a}}
\put(73.00,40.33){\vector(0,-1){0.2}}
\emline{73.00}{45.00}{7}{73.00}{40.33}{8}
\put(73.00,34.00){\vector(0,1){0.2}}
\emline{73.00}{30.67}{9}{73.00}{34.00}{10}
\end{picture}
\end{center}
\caption{\label{nf1} An infinite crystalline atomic plane
perpendicular to the paper, and an incident plane wave with wave
vector ~$\kk_0$. Atoms are in the plane $x,y$, which crosses
$z$-axis at the point ${z=0}$. An elementary cell of the plane is
a square with sides of length $a$ along $x$ and $y$-axes. The
axis~$y$ is perpendicular to the figure plane and is directed
toward the reader.}
\end{figure}

Now if we substitute \eref{n8a} into \eref{n7a} we obtain the full
wave function with scattering on a crystalline plane to be
\begin{equation}
\psi(\rr)=\exp(i\kk_0\cdot\rr)
-bC\sum_{\sn}\exp(ia\kk_0\cdot\n)\fr{\exp(ik|\rr-a\n|)}{|\rr-a\n|},
\label{n7a1}\end{equation} where summation goes over infinite
number of atoms in the crystalline plane~\cite{glas}. It looks as
if we use perturbation theory without multiple scattering, and all
the multiple scattering is contained in the renormalization factor
$C$. We will evaluate and discuss this factor later, but want to
warn the reader here that it may happen that in rigorous standard
scattering theory $C=0$, i.e. the crystalline plane cannot scatter
at all and is invisible for the incident plane wave. We know that
the diffraction exists and will not believe such an extraordinary
possible result $C=0$.

\section{Diffraction from a single crystalline plane}

We found the total wave function \eref{n7a1} but it tells nothing.
We need to know how to make a summation over all the atoms.
However there is a simple recipe how to deal with arbitrary sums.

\subsection{A recipe for summation}

Let's consider a sum
\begin{equation}\label{n6a}
S=\sum\limits_{n=n_1}^{n_2}f(n)
\end{equation}
for an arbitrary function $f(n)$. This sum can be represented in a
different form
\begin{equation}\label{n6b}
S=\sum\limits_{N=-\infty}^{+\infty}F(N),
\end{equation}
where
\begin{equation}\label{n6c}
F(N)=\int\limits_{n_1}^{n_2}f(n)\exp(2\pi iNn)dn,
\end{equation}
and $n$ is considered as a continuous variable.

\begin{figure}[h!]
{\par\centering\resizebox*{12cm}{!}{\includegraphics{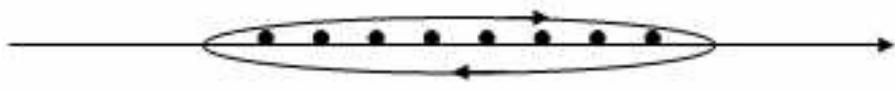}}\par}
\caption{A closed path for integration \eref{n6e} in the complex
plane.} \label{nf2}
\end{figure}

The recipe can be easily checked. One can apply the same
transformation to \eref{n6b}. As a result we obtain
\begin{equation}\label{n6d}
S=\sum\limits_{M=-\infty}^{+\infty}\int\limits_{-\infty}^{+\infty}dN\exp(2\pi
MN)\int\limits_{n_1}^{n_2}f(n)\exp(2\pi iNn)dn=$$
$$=\sum\limits_{M=-\infty}^{+\infty}\int\limits_{n_1}^{n_2}f(n)\exp(2\pi
iNn)dn\delta(2\pi(M+n))=\sum\limits_{n=n_1}^{n_2}f(n).
\end{equation}
It looks like a trick, but can be rigorously proven. Indeed, the
sum \eref{n6a} can be represented as an integral
\begin{equation}\label{n6e}
\sum\limits_{n=n_1}^{n_2}f(n)=\oint\limits_C
dz\fr{f(z)}{1-\exp(2\pi iz)}
\end{equation}
over closed path in complex plane as is shown in fig, \ref{nf2}.
Expansion of the function $1/(1-\exp(2\pi iz))$ over powers
$\exp(2\pi Niz)$ on upper half of the path, and over powers
$\exp(-2\pi Niz)$ on lower half of the path gives the sum
\eref{n6b} with definition \eref{n6c}.

It looks not profitable to transform a final sum into an infinite
one. However in our case of the infinite sum \eref{n7a1} it is
profitable. Generalization of \eref{n6c} to double sum is trivial,
so we can represent \eref{n7a1} in the form
\begin{equation}
\psi(\rr)=\exp(i\kk_0\cdot\rr) -bC\sum_{\sm}\int d^2n\exp(2\pi
i\m\n) \exp(ia\kk_0\cdot\n)\fr{\exp(ik|\rr-a\n|)}{|\rr-a\n|},
\label{n7a2}\end{equation} or
\begin{equation}
\psi(\rr)=\exp(i\kk_0\cdot\rr) -b_CN_2\sum_{\sta}\int
d^2r'_\|\exp(i(\ta+\kk_0)\cdot\rr'_\|)
\fr{\exp(ik|\rr-\rr'_\||)}{|\rr-\rr'_\||},
\label{n7a22}\end{equation} where $\rr'_\|$ are coordinates in the
plane, $N_2=1/a^2$ is atomic density on the plane, $b_C=bC$ is
renormalized scattering amplitude of a single atom, and we
introduced vectors $\ta=2\pi\m/a$ of the reciprocal plane lattice.
From \eref{n7a22} it follows that we need two-dimensional Fourier
expansion of the spherical waves. Let's find it.

\subsection{3-dimensional Fourier expansion of spherical waves}

To find two-dimensional expansion we start with the well known
3-dimensional Fourier expansion of the spherical wave. It looks
\begin{equation}\label{n10}
\fr{\exp(ikr)}{r}=\fr{4\pi}{(2\pi)^3}\int
d^3p\fr{\exp(i\p\rr)}{p^2-k^2-i\varepsilon}.
\end{equation}
To prove that the right side is equal to the left function, one
represents $d^3p=p^2dpd\varphi d\cos\theta$, integrates the right
hand side over angles and obtains
\begin{equation}\label{n11}
\fr{1}{i\pi r}\int\limits_0^\infty
pdp\fr{\exp(ipr)-\exp(-ipr)}{p^2-k^2-i\varepsilon}=\fr{1}{i\pi
r}\int\limits_{-\infty}^\infty
pdp\fr{\exp(ipr)}{p^2-k^2-i\varepsilon}.
\end{equation}
Since $r>0$ the integration path can be closed in complex plane of
the integration variable $p$ by the infinite semicircle in the
upper half part of the plane, and the result of the integration
over closed path is the residual in the single pole at
$p=k+i\epsilon$. This residual is $2\pi ik\exp(ikr)/2k$, and its
substitution into \eref{n11} gives the left hand side of
\eref{n10}.

From \eref{n10} it is easy to deduce what equation does spherical
wave satisfy. Indeed, if we apply to it the operator $\Delta+k^2$,
we obtain
\begin{equation}\label{n11a}
(\Delta+k^2)\fr{\exp(ikr)}{r}=\fr{4\pi}{(2\pi)^3}\int
d^3p\fr{(k^2-p^2)\exp(i\p\rr)}{p^2-k^2-i\varepsilon}=-4\pi\delta(\rr).
\end{equation}

\subsection{2-dimensional Fourier expansion of spherical waves}

However we need not the 3-dimensional but 2-dimensional Fourier
expansion of the spherical wave. It is obtained from \eref{n10} by
representation $d^3p=dp_zd^2p_\|$, and
$p^2-k^2-i\varepsilon=p_z^2-p_\bot^2-i\varepsilon$, where
$p_\bot=\sqrt{k^2-p^2_\|}$ and vector $\p_\|$ lies in the $(x,y)$
plane. As a result we obtain
\begin{equation}\label{n10a}
\fr{\exp(ikr)}{r}=\fr{4\pi}{(2\pi)^3}\int
d^2p_\|\int\limits_{-\infty}^\infty
dp_z\fr{\exp(i\p_\|\rr_\|+ip_zz)}{p_z^2-p_\bot^2-i\varepsilon}.
\end{equation}
The integrand has two poles at $p_z=\pm (p_\bot+i\varepsilon)$.
The integration path can be closed in complex plane of the
integration variable $p_z$ by an infinite semicircle in the upper
half part of the plane, when $z>0$, and in the lower half part of
the plane, when $z<0$. In both cases inside the closed path there
is only one pole, so the result of the integration is
\begin{equation}\label{n10b}
\fr{\exp(ikr)}{r}=\fr{i}{2\pi}\int
\fr{d^2p_\|}{p_\bot}\exp(i\p_\|\rr_\|+ip_\bot|z|).
\end{equation}
One can directly integrate the right hand side to get spherical
wave, if for every $r$ one directs $z$-axis toward $r$ so that
$\rr_\|=0$, and then the integral in the right hand side of
\eref{n10b} is easily calculated.

\subsection{Digression on contradictions in quantum scattering
theory}

Let's look at the wave function \eref{n1}. It contains the
incident wave. which satisfies free equation,
\begin{equation}\label{n12}
(\Delta+k^2)\exp(i\kk_0\rr)=0,
\end{equation}
and the scattered spherical wave, which satisfies \eref{n12}. The
last one is not free. It is inhomogeneous. So the spherical wave
does not correspond to a free particle and should not be used. The
standard objection to this claim is: we do not worry about the
point $r=0$, and outside this point the spherical wave satisfies
the free \Sh. Therefore the spherical wave describes the free
particle.

\begin{figure}[h!]
{\par\centering\resizebox*{8cm}{!}{\includegraphics{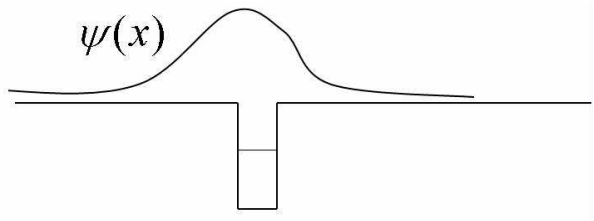}}\par}
\caption{A particle in a bound state in a potential well has
nonzero wave function outside of the well. The tails of this wave
function satisfy the free \Sh\ $(\Delta+k^2)\psi=0$. If potential
well is one dimensional like a potential trough, the motion along
the trough can be arbitrary, so the total energy $k^2$ of the
particle in \Sh\ outside of the well can be positive.}
\label{nf2a}
\end{figure}

However this argument is not appropriate. Indeed, let's consider a
potential well and a particle in a bound state in this well as
shown in fig. \ref{nf2a}. Outside of the well the particle
satisfies the free \Sh, however it is not free, and a
distinguishing feature of the bound state is exponential decay of
the wave function away from the well. In the case of a simple
spherical potential the kinetic energy $k^2$ in the \Sh\ outside
the well is negative. But we can imagine a cylindrical potential
well with arbitrary high movement along the cylinder. In that case
the total kinetic energy $k^2$ outside the well can be positive,
nevertheless the wave function exponentially decays away from the
potential.

The spherical wave, according to expansion \eref{n10b}, contains
exponentially decaying part. It is the part of the integral with
$p_\|>k$. If one excludes this part of the spherical wave, then
the remaining integral will be
\begin{equation}\label{n10c}
\fr{\exp(ikr)}{r}\Longrightarrow\fr{i}{2\pi}\int\limits_{p_\|<k}
\fr{d^2p_\|}{p_\bot}\exp(i\p_\|\rr_\|+ip_\bot|z|)=\fr{i}{\pi}\int\limits_{\p\cdot\rr>0}
d^3p\delta(p^2-k^2)\exp(i\p\cdot\rr),
\end{equation}
where integration limit warrants that the integral contains only
outgoing plane waves, so it cannot be reduced to imaginary part of
the spherical function
\begin{equation}\label{n10d}
i\fr{\sin(kr)}{r},
\end{equation}
which contains ingoing waves also.

With account of \eref{n10c} the scattered waves after integration
over $dp$ can be represented as
\begin{equation}\label{n10e}
\psi_{sc}(\rr)=-\fr{ikb}{2\pi}\int\limits_{\kk_\Omega\cdot\rr>0}
d\Omega\exp(i\kk_\Omega\cdot\rr),
\end{equation}
where vector $\kk_\Omega$ has length $k$ and direction determined
by the solid angle $\Omega$. From this expression it follows, that
probability of scattering in the direction $\OO$ is equal to
\begin{equation}\label{n10f}
dw(\Omega)=|b/\lambda|^2d\Omega.
\end{equation}
It is dimensionless, and such a dimensional parameter as a cross
section can be defined only artificially. To have a consistent
theory one needs to work with wave packets and to introduce
nonlinearity. Wave packet without nonlinear wave equation does not
help.

In our diffraction problems we will deal only with probabilities,
so no problem with definition of cross sections will arise.

\subsection{Diffraction on the crystalline plane}

Let's substitute \eref{n10b} into \eref{n7a22}, then, having in
mind that all atoms in our crystalline plane have $z_n=0$ we get
\begin{equation}\label{na7}
\psi(\rr)=\exp(i\kk_0\cdot\rr)
-\fr{iN_{2}b_c}{2\pi}\sum_{\sta}\int
d^2r'_\|\exp(i(\ta+\kk_0)\cdot\rr'_\|)
\int\fr{d^2p_\|}{p_\bot}\exp(i\p_\|\cdot(\rr_\|-\rr'_\|))=
\end{equation}
\begin{equation}
= \exp(i\kk_0\cdot\rr) -\sum_{\sta}
\fr{i\kappa}{k_{\tau\bot}}\exp(i\kk_{\tau\|}\cdot\rr_\|+ik_{\tau\bot}|z|),
\label{n7a3}\end{equation} where $\kk_{\tau\|}=\kk_{0\|}+\ta$,
$k_{\tau\bot}=\sqrt{k^2-\kk_{\tau\|}^2}$ and $\kappa=2\pi N_2b_C$.
Let's note that for thermal neutrons $\kappa/k_{\tau\bot}\approx
b\lambda/a^2$ is of the order $10^{-4}\ll1$, i.e. scattering on a
single crystalline plane is very small.

Now we can sum up. We see that scattering creates a set of
discrete diffracted plane waves going symmetrically on both sides
of the crystalline plane. The diffracted waves propagate with wave
vectors $\kk_{\sta}=(\kk_{\sta\|},k_{\sta\bot})$, where
$\kk_{\sta\|}=\kk_{0\|}+\ta$,
$k_{\sta\bot}=\sqrt{k^2-\kk_{\sta\|}^2}$, $\ta=\tau_1\n$,
$\tau_1=2\pi/a$, and $\n$ is a 2-dimensional vector with integer
components. The amplitudes of the waves are equal to
\begin{equation}\label{n13}
f_\tau=\fr{2\pi iN_2bC}{k_{\sta\bot}}.
\end{equation}
It seems that the number of diffracted waves is infinite, however
the real number of propagating plane waves is finite, because for
sufficiently large $\n$ the normal component of the wave vector
$k_{\sta\bot}$ becomes imaginary, and corresponding diffracted
waves exponentially decay away from the crystalline plane.
According to our consideration of the spherical wave we must
exclude exponentially decaying waves from the integral in
\eref{na7}, then the exponentially decaying waves will not appear
in \eref{n7a3} either. However sometimes we need exponentially
decaying waves. If near our crystalline plane there is another
plane, then an exponentially decaying wave from the first plane
can reach the second one and in the process of diffraction it will
create diffracted propagating waves. We will show it later.

It is worth also to discuss the amplitude \eref{n13} of the
diffracted waves. Since wave vector $\kk_0$ of the incident wave
can be arbitrary, it may happen that for some vector $\ta$ of the
reciprocal lattice denominator $k_{\sta\bot}$ of \eref{n13}
becomes so small that $|f_\tau|\gg1$. The question arises: what
does it mean, and whether it is really possible? To answer this
question it is now necessary to consider the role of the
renormalization factor $C$, which is not a constant but depends on
the incident wave vector $\kk_0$. The role of $C$ is to guard
unitarity, which is a requirement equivalent to the law of energy
conservation.

\subsection{Unitarity for a single scatterer}

First of all let's look at the simplest scattering wave function
given by \eref{n1}. It is easy to prove that to satisfy unitarity
in absence of absorption the scattering amplitude must be of the
form
\begin{equation}\label{n14}
b=\fr{b_0}{1+ikb_0},
\end{equation}
where $b_0$ is a real number. Let's prove it. Substitute
\eref{n10b} into \eref{n1}, and choose $z$-axis in the direction
of the incident wave propagation. Then the wave function becomes
\begin{equation}\label{n15}
\psi(\rr)=\exp(ikz)-\fr{ib}{2\pi}\int
\fr{d^2p_\|}{p_\bot}\exp(i\p_\|\rr_\|+ip_\bot|z|).
\end{equation}
Let's choose two planes perpendicular to $z$-axis before, $S_1$,
and behind, $S_2$, the scatterer. Unitarity means a requirement
that flux density $\ola{J}_{S_1}$ of particles going after
scattering to the left, plus the flux density $\ora{J}_{S_2}$ of
particles going after scattering to the right should be equal to
the flux density $\ora{J}_{0S_1}$ of the incident particles going
toward the scattering center through the plane $S_1$. Let's
calculate these fluxes.

\subsubsection{The incident flux density}
First of all let's remind definition of the flux density. For the
wave function $\psi(\rr)$ the flux density through a plane $S$
with the normal along $z$-axis is
\begin{equation}\label{n16}
\ora J=\lim_{S\to
\infty}\fr1{2iS}\int\limits_Sd^2x_\|\lt[\psi^*(\rr)\lt(\fr{\ora
d}{dz}-\fr{\ola d}{dz}\rt)\psi(\rr)\rt]_{z=0},
\end{equation}
where arrows over derivatives show which side should be
differentiated. So, for $\psi_0=\exp(ikz)$ the incident flux is
$\ora J_0=k$.
\subsubsection{Scattered flux density}
Substitution of the wave function \eref{n15} into \eref{n16} gives
the flux of the waves scattered to the left
\begin{equation}\label{n17}
\ola J_{sc}=\lt(\fr b{2\pi}\rt)^2\int\limits_{p_\|<k}
\fr{d^2p_\|}{p_\bot}\int\limits_{p'_\|<k}
\fr{d^2p'_\|}{p'_\bot}\int
d^2r\fr{p'_\bot+p_\bot}{2S}\exp(i(\p_\|-\p'_\|)\rr_\|)=\fr{2\pi|b|^2}{S}.
\end{equation}
Since the scattering is symmetrical, the same flux will be
obtained for the waves scattered to the right.

\subsubsection{Interference flux density}

Let's look carefully for the flux of particles going to the right
plane $S_2$ behind the scattering center. The wave function there
is a superposition of the incident and scattered waves
$\psi=\psi_0+\psi_{sc}$. Therefore, substitution of it into
\eref{n16} gives $\ora J_0$, $\ora J_{sc}$ and the interference
flux
\begin{equation}\label{n18}
\ora
J-{int}=\fr1{2iS}\int\limits_Sd^2x_\|\lt\{\lt[\psi_0^*(\rr)\lt(\fr{\ora
d}{dz}-\fr{\ola
d}{dz}\rt)\psi_{sc}(\rr)\rt]_{z=0}+\lt[\psi_{sc}^*(\rr)\lt(\fr{\ora
d}{dz}-\fr{\ola d}{dz}\rt)\psi_{0}(\rr)\rt]_{z=0}\rt\}=$$
$$=-2\pi i\fr{b-b^*}{S}.
\end{equation}
Let's sum up. Requirement of unitarity is
\begin{equation}\label{n19}
\ola J_{sc}+\ora J_{sc}+\ora J_0+\ora J_{int}=\ora J_0,
\end{equation}
from which it follows
\begin{equation}\label{n20}
\ola J_{sc}+\ora J_{sc} +\ora J_{int}=0,
\end{equation}
or
\begin{equation}\label{n20a}
4\pi|b|^2k+4\pi Im(b)=0.
\end{equation}
Finally we get the relation known as the optical theorem
\begin{equation}\label{n21}
Im(b)=-\fr{k\sigma}{4\pi}.
\end{equation}
where $\sigma=4\pi|b|^2$ --- cross section of elastic scattering.
Now, if we look at \eref{n14}, we see that it precisely satisfies
optical theorem \eref{n21}. We want also to add that removal of
exponentially decaying part of the spherical wave does not spoil
unitarity.

\subsection{Unitarity for scattering on a single crystalline plane}

Unitarity for crystalline plane is formulated with the same
\eref{n20}, however flux density with respect to the plane is
determined by the normal component of the wave vector. For
instance the flux of the incident wave $\ora J_0=k_{0\bot}$.

From \eref{n7a3} it follows that
\begin{equation}\label{n21a}
\ola J_{sc}=\ora J_{sc}=\sum_{\sta} \fr{|\kappa|^2}{k_{\tau\bot}},
\end{equation}
and
\begin{equation}\label{n21b}
\ora J_{int}=i(\kappa^*-\kappa).
\end{equation} Therefore, the unitarity condition means
\begin{equation}\label{n21c}
Im(\kappa)=-\sum_{\sta} \fr{|\kappa|^2}{k_{\tau\bot}},
\end{equation}
or
\begin{equation}\label{n21d}
Im(bC)=2\pi N_2|bC|^2\sum_{\sta} \fr{1}{k_{\tau\bot}}.
\end{equation}
Let's note, that $k_{\tau\bot}$ in sum are all real. Therefore
exponentially decaying waves do not contribute to the unitarity.

\subsubsection{Calculation of the factor $C$}

\begin{figure}[h!]
{\par\centering\resizebox*{8cm}{!}{\includegraphics{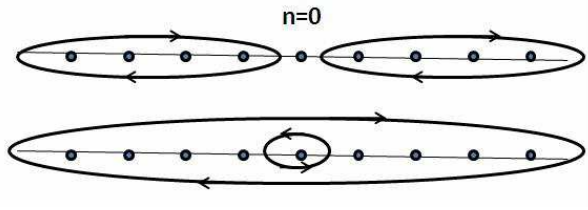}}\par}
\caption{To exclude one point $n=0$ in summation, one can use two
integrals of the type \eref{n6e} over two closed loops as shown in
the upper part of the figure. However it can be done as shown in
lower part. It is possible to make the single integral \eref{n6e}
over large closed path without exclusion and to add an integral
over closed loop around excluded point in the opposite direction.}
\label{nf3}
\end{figure}

 Now it is the time to calculate renormalization factor $C$ to prove that it really helps to satisfy
 \eref{n21d}. To
do that we can apply to \eref{n9} the sum rule
\eref{n6b}-(\ref{n7a22}). If there were not limitation $n\ne0$ in
\eref{n9}, we would obtain directly like in \eref{n7a3}
$$S=\sum_{\sn}\exp(i(\kk_0\cdot\n)
a)\fr{\exp(ika|\n|)}{a|\n|}=N_2\sum_{\sta}\int
d^2r'_\|\exp(i(\ta+\kk_0)\cdot\rr'_\|)
\fr{\exp(ik|\rr'_\||)}{|\rr'_\||}=$$
\begin{equation}
=\fr{iN_{2}}{2\pi}\sum_{\sta}\int
d^2r'_\|\exp(i(\ta+\kk_0)\cdot\rr'_\|)
\int\fr{d^2p_\|}{p_\bot}\exp(i\p_\|\cdot(\rr_\|-\rr'_\|))=\sum_{\sta}
\fr{2\pi iN_2}{k_{\tau\bot}}. \label{n9a}\end{equation} However
the term with $\n=0$ in the left hand side is singular and
unacceptable. Therefore it must be excluded. Exclusion can be made
as shown in fig. \ref{nf3}. The integral over large closed loop
gives the result in the right hand side of \eref{n9a}, and to
exclude the point $n=0$ one adds to \eref{n9a} the counter
clockwise integral around this point along a small circle, which
gives
\begin{equation}\label{n22}
I_0=\oint \fr{dn}{1-\exp(2\pi in)}\fr{\exp(ikan)}{an}=-\oint
\fr{dn}{2\pi ian^2}\exp(ikan)=-ik.
\end{equation}

Let's note, that if we do not remove from the spherical wave the
exponentially decaying part, then the sum in \eref{n9a} contains
real and imaginary $k_{\tau\bot}$. So, according to \eref{n9},
(\ref{n9a}), (\ref{n22}) and \eref{n14} the renormalized amplitude
is
\begin{equation}
b_C=bC=\fr{b}{1+b\lt(-ik+\sum_{\sta} \fr{2\pi
iN_2}{k_{\tau\bot}}\rt)}=\fr{b_0}{1+ikb_0-ikb_0+b_0S'+ib_0S''}=\fr{b_1}{1+ib_1S''},
\label{n9b}\end{equation} where
\begin{equation}\label{n9c}
S'=\sum_{k_{\tau\|}>k} \fr{2\pi N_2}{|k_{\tau\bot}|},\qquad
S''=\sum_{k_{\tau\|}<k} \fr{2\pi N_2}{k_{\tau\bot}},\qquad
b_1=\fr{b_0}{1+b_0S'}.
\end{equation}
We see that normalization factor first of all cancels the term
$ikb_0$ in \eref{n14}, which provides unitarity for a single
scatterer, next, it changes the real value $b_0$ to $b_1$ and
finally provides new imaginary part $ib_1S''$ in denominator,
which is appropriate to satisfy unitarity for the crystalline
plane. It is trivial to see that \eref{n9b} does satisfy relation
\eref{n21d}.

\subsubsection{A problem of renormalization of the real part
$b_0$ in \eref{n9c}}

The real part $b_0$ of the amplitude $b$ is renormalized, as
follows from \eref{n9c}, only with that part of the sum, which
contains imaginary $k_{\tau\bot}$. It stems from exponentially
decaying part of the spherical wave. The sum
\begin{equation}\label{n23}
S'=\sum_{k_{\tau\|}>k} \fr{2\pi N_2}{|k_{\tau\bot}|},
\end{equation}
for large $k_{\tau\|}=|\kk_{0\|}+(2\pi/a)\n|$, where
$\n=(n_x,n_y)$ has large integer components $n_{x,y}$, can be
approximated as
\begin{equation}\label{n23a}
S'=2\pi N_2a\sum_{n_x,n_y} \fr{1}{\sqrt{n_x^2+n_y^2}},
\end{equation}
and such a sum is diverging. Therefore the renormalized value
$b_1$ becomes zero, the crystalline plane becomes invisible to
incident waves and, contrary to our experience, creates no
diffraction. It shows once again how contradictory is description
of scattering with the help of spherical waves. If we exclude
exponentially decaying part from spherical wave, the real part
$b_0$ of the scattering amplitude will not be renormalized.

\subsubsection{Solution for singularity at $k_{\tau\bot}=0$ in \eref{n7a3}}

Now we can resolve the problem, which appears in \eref{n7a3} for
some wave vectors $\kk_0$ of the incident wave, for which one of
$k_{\tau\bot}$ is close to zero. In that case the factor $C$ goes
to zero too because it contains $1/k_{\tau\bot}$ in its
denominator. Therefore the amplitudes of all the diffracted waves
go to zero except the wave, which amplitude contains the same
factor $1/k_{\tau\bot}$. For this wave the divergent factors
cancel, and the diffraction gives a single diffracted wave with
unit amplitude, which propagates along the crystalline plane. It
would be very interesting to observe such an effect, which can be
seen not only for a single crystalline plane, but also for the
whole crystal.

\section{Conclusion}

The story about crystalline plane and a scalar wave is over.
However this story is only a beginning of many other stories,
where one can deduce optical potential of media, diffraction of
scalar waves on single crystals, diffraction of electromagnetic
waves and many others, some of which possibly will be published in
this journal. However one of the most important point of the
finished story is the pinpointed contradiction of scattering
theory. Without resolution or discussion of it quantum theory is
doomed to stagnation.

\section*{Acknowledgements}
I am thankful to my students, who were patient, polite, and their
eyes some times gave me an impression that they enjoyed my
lectures. I am also grateful to the single referee of Am.J.Phys.,
who appreciated my paper. He is great, but his voice was drowned
in between voices of those, who can only learn, but are not able
to think.


\begin{thebibliography}{9}
\bibitem{ew}
Paul Peter Ewald ``Kristalle und Rжntgenstrahlen'' (Springer,
1923)
\bibitem{bac}
G.E. Bacon ``Neutron Diffraction'' Oxford University Press 1955.
\bibitem{bush}
R.T. Bush ''The Ewald Construction and the Bragg Law of
Diffraction in a First Course in Solid-State Physic.'' Am. J.
Phys. 37, 669 (1969)
\bibitem{hal}
B. D. Hall ``Introductory crystallography for solid-state physics:
Computer simulations in two dimensions.'' Am. J. Phys. 66, 19
(1998).
\bibitem{uig} M. Utsuro, V. K.  Ignatovich,
\textit{ Handbook of Neutron Optics}  (Wiley-VCH, Weinheim, 2010).
\bibitem{conc1}
Ignatovich VK. Contradictions in scattering theory. Concepts of
Phys.2004;1:51-104.
\bibitem{fold}
Foldy L.L. ``The multiple scatteling of waves.'' Phys.Rev.
{\bf67}(3) 107-19 (1945)
\bibitem{lax}
Lax M. ``Multiple scattering of waves''.
Rev.Mod.Phys.{\bf23}287-310 (1951).
\bibitem{lax1}
Lax M. ``Multiple scattering of waves II. The Effective Field in
Dense Systems''. Phys.Rev. {\bf 85}(4) 621-629 (1952).
\bibitem{glas}
Glasser M.L. ``The evaluation of lattice sums. I. Analytic
procedures.'' J.Math.Phys.{\bf14}(3) 409-413 (1973).
\end{thebibliography}
\end{document}